\documentclass[%
 reprint,
 amsmath,amssymb,
 aps,
]{revtex4-2}

\usepackage{graphicx}
\usepackage{dcolumn}
\usepackage{bm}
\usepackage{hyperref}
\usepackage[mathlines]{lineno}
\usepackage{graphics}
\usepackage{multirow}
\usepackage{tikz}
\usepackage{amssymb}
\usepackage{amsmath}

\begin{document}
 
\preprint{APS/123-QED}

\title{Ground and Excited state masses of $\Omega_c^0$, $\Omega_{cc}^+$ and $\Omega_{ccc}^{++}$ baryons}

\author{Juhi Oudichhya}
\author{Keval Gandhi}%
 \email{keval.physics@yahoo.com}
 \author{Ajay Kumar Rai} 
\affiliation{Department of Physics, Sardar Vallabhbhai National Institute of Technology, Surat, Gujarat-395007, India}

%
%

\date{\today}

\begin{abstract}
The masses of ground and excited states of $\Omega_c^0$, $\Omega_{cc}^+$ and $\Omega_{ccc}^{++}$ baryons are calculated within the framework of Regge phenomenology. Here our attempt is to assign a possible spin-parity to these obtained masses. Our calculated masses in $(J,M^2)$ plane with both natural ($J^{P}$ = $1/2^{+}$, $3/2^{-}$, $5/2^{+}$, ....) and unnatural ($J^{P}$ = $3/2^{+}$, $5/2^{-}$, $7/2^{+}$, ....) spin-parities are in agreement with the experimental observations where available and reasonably close to other theoretical predictions. Further we fix the slope and intercept of each Regge line in $(n, M^2)$ plane and estimate the masses of higher excited states of these baryons. The obtained mass relations and the mass value predictions could provide an useful information in future experimental searches and the spin-parity assignment of these states. 
\end{abstract}

\maketitle


\section{\label{sec:level1} Introduction}  
In 2017, the LHCb collaboration observed five new narrow excited $\Omega_{c}^{0}$ states such as $\Omega_{c}(3000)^{0}$, $\Omega_{c}{(3050)}^0$, $\Omega_{c}{(3066)}^0$, $\Omega_{c}{(3090)}^0$ and $\Omega_{c}{(3119)}^0$ decaying into $\Xi^{+}K^{-} $ \cite{LHCb}. Except for $\Omega_{c}{(3119)}^0$ other four states were further confirmed by Belle \cite{Belle} in 2018. Frequent questions have been arises with this discovery, for instance \textit{(i) why only five states were observed ? (ii) why they are so narrow ?}. Many authors have been addressed these questions with recent studies given in Refs. \cite{Rosner,M.Padmanath2017,B.Chen2017,Wang2018,ZWang2017,Agaev2017,HChen2017,HCheng2017,KWang2017,Zhao2017,Yang2018,Haung2018,C.S2017}. Since $\Omega_{c}^{0}$ is composed of two strange quarks $(ss)$ and one charm quark $(c)$, if $ss$ diquark remains in its color triplet it has spin-1, as symmetry rules do not allow a spin-0 diquark. This spin combined with spin 1/2 of $c$ quark, we have $S = 1/2$ or $3/2$. Now consider states with relative orbital momentum $L=1$ and combining with $S = 1/2$ and $3/2$ gives the total spin $J = 1/2$, $3/2$ and $J = 1/2$, $3/2$, $5/2$, respectively. All these five states have negative parity. Also these states are extremely narrow which indicates that it is difficult to pull apart the two $s$ quarks in a diquark \cite{Rosner}. 

The study of heavy-light bound state provides an ideal platform to understand the dynamics of quantum chromodynamics (QCD) at low energy regime. The strong interaction coupling is small for the hadrons containing heavy quark ($c$ or $b$) which makes it easier to understand the QCD perturbatively than the system containing only light quark \cite{Manoharbook}. Different experimental facilities have provided new information in the sector of hadron spectroscopy like masses, decay width, branching ratios, isospin mass splittings, apin, parity, polarisation amplitude, etc. It is very crucial to assign the spin-parity of hadrons which facilitate the determination of experimental properties. Experimentally only the ground state $\Omega_{c}^{0}$ and $\Omega_{c}(2770)^{0}$ are observed with their quantum numbers, $J^{P}$ = $\frac{1}{2}^{+}$ and $\frac{3}{2}^{+}$ respectively \cite{PDG}, where $J$ is the total spin and $P$ denotes the parity. The $J^{P}$ values of excited state $\Omega_{c}^{0}$ baryons are still missing as shown in Table \ref{tab:tablea}. 

Upto the date there are different research aspects to study the hadron spectroscopy, one of the prominent way to determine the properties of hadronic states is mass spectrum analysis. So far there are various method to depict the mass spectrum. The authors of Refs. \cite{ZalakC2016,Zalak2017,keval2020,Zalak2016} calculate the excited state masses of singly, doubly and triply heavy baryons using non-relativistic approach of hypercentral constituent quark model (hCQM). In Ref. \cite{Djia2020} mass analysis is carried out for the newly excited $\Omega_{c/b}$ baryons by proposing a new scheme of state classification called $Jls$ mixing coupling which goes beyond $jj$ coupling. The results favor to predict the spin-parity of these states. The ground state and excited state spectra of $\Omega_{c}^{0}$ baryon are estimated from lattice QCD with dynamical quark field in Ref. \cite{M.Padmanath2017} and results predicted the $J^{P}$ assignments of the states $\Omega_{c}(3000)^{0}$, $\Omega_{c}{(3050)}^0$ and $\Omega_{c}{(3066)}^0$, $\Omega_{c}{(3090)}^0$ and $\Omega_{c}{(3119)}^0$ with $1/2^{-}$, $3/2^{-}$ and $5/2^{-}$, respectively. The authors of Ref. \cite{Rosner} predicts the existence of five states with negative parity by assuming these baryonic states as bound state of $c$-quark and a $P$ wave $ss$-diquark. Ebert et al. \cite{Ebert 2011} determine mass spectra of excited heavy baryons consisting of two light $(u,d,s)$ and one heavy $(c,b)$ quarks, in the framework of QCD-motivated by relativistic quark model. Robert et al.\cite {Robert2008} used non-relativistic hamiltonian quark model to calculate the masses of heavy baryons. Ref. \cite{T.M 2020} estimate the masses and residues of these newly observed $\Omega_{c}^{0}$ baryon states within QCD sum rules and try to determine the nature of these states. So far none of the doubly/triply $\Omega$ baryons have been experimently detected. Although many authors have predicted the masses of $\Omega_{cc}$ and $\Omega_{ccc}$ baryons using various theoretical and phenomenological models, like Relativistic quark model \cite{A.P2008} , Salpeter model \cite{F.G2009}, Lattice QCD \cite{CAlex2014y,Brown2014}, the Hamiltonian model \cite{Yoshida2015}, the Hypercentral constituent quark model \cite{ZalakC2016,Zalak2017} etc.

Wei et al. \cite{Wei2008} derive some useful mass relations such as linear mass inequalities, quadratic mass inequalities and quadratic mass equalities for hadrons using quasilinear Regge trejectory ansatz. Based on these relations they determine mass ranges of some mesons and baryons and predict $J^{P}$ assignments of $\Xi_{c}$(2980), $\Xi_{c}$(3055), $\Xi_{c}$(3077) and $\Xi_{c}$(3123) baryons. Ref. \cite{Wei2015} the same group extended this methodology to compute the masses of ground state and excited state of other doubly and triply charmed baryons such as $\Omega_{cc}, \Xi_{cc}^{*}, \Omega_{cc}^{*}$ and $\Omega_{ccc}$. In the present work we have used the same approach of Regge phenomenology with the assumption of linear Regge trejectories. We obtained relations between  intercept, slope ratios and baryon masses in both ($J,M^{2}$) and ($n_{r},M^{2}$) planes. Using these relations we determine the mass spectra of $\Omega_{c}$ baryon and try to assign a possible spin-parity to these recently observed five baryonic states. Further we apply the same scheme  for doubly and triply charmed omega baryons. Since these baryons are yet to be observed by experiments and therefore they are not listed in PDG \cite{PDG} yet. This motivate us to study the mass spectra of $\Omega_{c}^{0}$, $\Omega_{cc}^{+}$ and $\Omega_{ccc}^{++}$ baryons. \\

\begin{table}
	\caption{\label{tab:tablea}
	Masses, width and $J^{P}$ value of $\Omega_{c}^{0}$ baryons reported by LHCb \cite{LHCb}.
	}
	\begin{ruledtabular}
		\begin{tabular}{llll}
			Resonance & Mass(MeV) & Width(MeV) & $J^{P}$  \\ 
			\colrule

			$\Omega_{c}^{0}$ & 2695.2$\pm1.7$ & - & $\frac{1}{2}^{+}$ \\
			$\Omega_{c}(2770)^{0}$ & 2765.9$\pm2.0$& - &$\frac{3}{2}^{+}$  \\
			$\Omega_{c}(3000)^{0}$ & 3000.4$\pm0.22\pm0.1$& $4.5\pm0.6\pm0.3$ &$?^{?}$  \\
			$\Omega_{c}(3050)^{0}$ & 3050.2$\pm0.1\pm0.1$& $0.8\pm0.2\pm0.1$ &$?^{?}$  \\
			$\Omega_{c}(3065)^{0}$ & 3065.6$\pm0.1\pm0.3$& $3.5\pm0.4\pm0.2$ &$?^{?}$ \\
			$\Omega_{c}(3090)^{0}$ & 3090.2$\pm0.3\pm0.9$& 8.7$\pm0.1\pm0.8$ &$?^{?}$ \\
			$\Omega_{c}(3120)^{0}$ & 3119$\pm0.3\pm0.9$& 1.1$\pm0.8\pm0.4$ &$?^{?}$\\
		\end{tabular}
	\end{ruledtabular}
\end{table} 
The remainder of this paper is organised as follows. After introduction, in sec II  we briefly explain the linear Regge trejectories and extract the quadratic mass relation in $(J,M^{2})$ plane to calculate the ground state and excited state masses of $\Omega_{c}$ baryon. Further we extend this work and again derive some mass equations to evaluate the mass spectra for $\Omega_{cc}$ and $\Omega_{ccc}$ baryons. In addition, we estimate the masses of these baryons in $(n,M^{2})$ plane by calculating Regge slopes and intercepts of particular regge lines. The discussions of our results appear in sec.III.
\section{\label{sec:level1}Theoretical Framework} 
Regge theory is one of the simplest and effective phenomenological approach to study the hadron spectroscopy. Various theories were developed to understand the regge trejectory. Among them, one of the straightforward explaination for linear regge trejectories were proposed by Nambu \cite{Nambu1974,Nambu1979}. He assumed that the uniform interaction of quark and antiquark pair form a strong flux tube and at the end of the tube the light quarks rotating with the speed of light at radius $R$. The mass originating in this flux tube is estimated as \cite{chen2017},
\begin{equation}
	\label{eq:1}
	M = 2\int_{0}^{R}\dfrac{\sigma}{\sqrt{1-\nu^{2}(r)}}dr = \pi\sigma R ,
\end{equation}
where $\sigma$ represent the string tension, i.e the mass density per unit length. Likewise, the angular momentum of this flux tube is evaluated as,
\begin{equation}
	\label{eq:2}
	J = 2\int_{0}^{R}\dfrac{\sigma r \nu(r)}{\sqrt{1-\nu^{2}(r)}}dr = \dfrac{\pi\sigma R^{2}}{2}+c^{'} ,
\end{equation}
one can also write,
\begin{equation}
	\label{eq:3}
	J = \dfrac{M^{2}}{2\pi\sigma} + c^{''} ,
\end{equation}
where $c^{'}$ and $c^{''}$ are the constants of integration. Hence we can say that, $J$ and $M^{2}$ are linearly related to each other.
The plots of Regge trejectories of hadrons in the $(J,M^{2})$ plane are usually called Chew-frautschi plots \cite{Chew1961}. Also hadrons lying on the same Regge line possess the same internal quantum numbers. From Eq. (\ref{eq:3}), the most general form of linear regge trejectories can be expressed as \cite{Wei2008},
\begin{equation}
	\label{eq:4}
	J = \alpha(M) = a(0)+\alpha^{'} M^2 ,
\end{equation}
where $a(0)$ and $\alpha^{'}$ represents the intercept and slope of the trejectory respectively. These parameters for different quark constituents of a baryon multiplet can be related by two relations \cite{Wei2008,Add1,Add2,Add3,Add4},
\begin{equation}
	\label{eq:5}
	a_{iiq}(0) + a_{jjq}(0) = 2a_{ijq}(0) ,	
\end{equation}

\begin{equation}
	\label{eq:6}
	\dfrac{1}{{\alpha^{'}}_{iiq}} + \dfrac{1}{{\alpha^{'}}_{jjq}} = \dfrac{2}{{\alpha^{'}}_{ijq}} ,	
\end{equation}\\
where $i, j, q$ represent quark flavors, $m_{i}<m_{j}$ and $q$ denotes an arbitrary light or heavy quark. Using Eqs. (\ref{eq:4}) and (\ref{eq:5}) we obtain,
\begin{equation}
	\label{eq:7}
	\alpha^{'}_{iiq}M^{2}_{iiq}+\alpha^{'}_{jjq}M^{2}_{jjq}=2\alpha^{'}_{ijq}M^{2}_{ijq} .
\end{equation}

Combining the relations (\ref{eq:6}) and (\ref{eq:7}) and after solving the quadratic equation, we obtain a pair of solutions as,
\begin{widetext}
	\begin{equation}
		\label{eq:8}
		\dfrac{\alpha^{'}_{jjq}}{\alpha^{'}_{iiq}}=\dfrac{1}{2M^{2}_{jjq}}\times[(4M^{2}_{ijq}-M^{2}_{iiq}-M^{2}_{jjq})
		\pm\sqrt{{{(4M^{2}_{ijq}-M^{2}_{iiq}-M^{2}_{jjq}})^2}-4M^{2}_{iiq}M^{2}_{jjq}}].
\end{equation}
\end{widetext}

This is the important relation we  obtained between slope ratios and baryon masses. Eq. (\ref{eq:8}) can also be expressed in terms of baryon masses by introducing $k$, where $k$ can be any quark flavor,	
\begin{equation}
	\label{eq:9}
	\dfrac{\alpha^{'}_{jjq}}{\alpha^{'}_{iiq}}=	\dfrac{\alpha^{'}_{kkq}}{\alpha^{'}_{iiq}}\times	\dfrac{\alpha^{'}_{jjq}}{\alpha^{'}_{kkq}} ,
\end{equation}	
we have,
\begin{widetext}
	\begin{equation}
		\label{eq:10}
		\begin{split}
			\dfrac{[(4M^{2}_{ijq}-M^{2}_{iiq}-M^{2}_{jjq})+\sqrt{{{(4M^{2}_{ijq}-M^{2}_{iiq}-M^{2}_{jjq}})^2}-4M^{2}_{iiq}M^{2}_{jjq}}]}{2M^{2}_{jjq}} \\
		=\dfrac{[(4M^{2}_{ikq}-M^{2}_{iiq}-M^{2}_{kkq})+\sqrt{{{(4M^{2}_{ikq}-M^{2}_{iiq}-M^{2}_{kkq}})^2}-4M^{2}_{iiq}M^{2}_{kkq}}]/2M^{2}_{kkq}}{[(4M^{2}_{jkq}-M^{2}_{jjq}-M^{2}_{kkq})+\sqrt{{{(4M^{2}_{jkq}-M^{2}_{jjq}-M^{2}_{kkq}})^2}-4M^{2}_{jjq}M^{2}_{kkq}}]/2M^{2}_{kkq}} .
		\end{split}
   \end{equation}
\end{widetext}

This is the general relation we have derived in terms of baryon masses which can be used to predict the mass of any baryon state by knowing all other masses. 

\subsection{Masses of $\Omega_{c}^{0}$ baryon}
In the present work we used Eq. (\ref{eq:10}) to evaluate the ground state masses of $\Omega_{c}^{0}$ baryon. Since $\Omega_{c}^{0}$ is composed of two strange quark and one charmed quark $(ssc)$. So keeping this configuration in mind we put $i=u$, $j=s$, $q=c$, $k=u$ in Eq. (\ref{eq:10}) and we get,

\begin{widetext}
	\begin{equation}
		\label{eq:11}
		\left[2M_{\Xi_{c}^{+}}(M_{\Omega_{c}^{0}}-2M_{\Xi_{c}^{+}})-M^{2}_{\Omega_{c}^{0}}-M^{2}_{\varSigma_{c}^{++}}\right] 
			= \sqrt{(4M^{2}_{\Xi_{c}^{+}}-M^{2}_{\varSigma_{c}^{++}}-M^{2}_{\Omega_{c}^{0}})^{2}-4M^{2}_{\Omega_{c}^{0}}M^{2}_{\Xi_{c}^{+}}} ,
				\end{equation}
	\end{widetext}
substituting the masses of $\Xi_{c}^{+}$ $(J^{P} = 1/2^{+}, 3/2^{+})$ and $\varSigma_{c}^{++}$ $(J^{P} = 1/2^{+}, 3/2^{+})$ baryon from PDG \cite{PDG} into relation (\ref{eq:11}), we obtain the ground state masses of $\Omega_{c}^{0}$ baryon such as 2.702 GeV for $J^{P} = \frac{1}{2}^{+}$ and 2.772 GeV for $J^{P} = \frac{3}{2}^{+}$. From Eq. (\ref{eq:8}) we can write , 
\begin{equation}
	\label{eq:12}
	\begin{split}
		\dfrac{\alpha^{'}_{ssc}}{\alpha^{'}_{uuc}}=\dfrac{1}{2M^{2}_{\Omega_{c}^{0}}}\times[(4M^{2}_{\Xi_{c}^{+}}-M^{2}_{\varSigma_{c}^{++}}-M^{2}_{\Omega_{c}^{0}}) \\  +\sqrt{{{(4M^{2}_{\Xi_{c}^{+}}-M^{2}_{\varSigma_{c}^{++}}-M^{2}_{\Omega_{c}^{0}}})^2}-4M^{2}_{\varSigma_{c}^{++}}M^{2}_{\Omega_{c}^{0}}}] .
	\end{split}
\end{equation}

Putting the values of masses in above equation, we get $\alpha^{'}_{ssc}/\alpha^{'}_{uuc}$. We can determine the slope by using Eq. (\ref{eq:4}),
\begin{equation}
	\alpha^{'} = \dfrac{(J+1)-J}{M^{2}_{J+1}-M^{2}_{J}} ,\nonumber
\end{equation}
for $\varSigma_{c}^{++}$$(uuc)$ it will be,
\begin{equation}
	\label{eq:13}
	\alpha^{'}_{uuc}=\dfrac{1}{M^2_{\varSigma^{++}_{c}(\frac{3}{2}^{-})}-M^2_{\varSigma^{++}_{c}(\frac{1}{2}^{+})}} ,
\end{equation}
we get $\alpha^{'}_{ssc} = $ 0.54835 $GeV^{-2}$ for $J^{P}=\frac{1}{2}^{+}$. Similarly we get  $\alpha^{'}_{ssc} = $ 0.66524 $GeV^{-2}$ for $J^{P}=\frac{3}{2}^{+}$. 

From Eq. (\ref{eq:4}) one can have,
\begin{equation}
	\label{eq:14}
	M_{J+1} = \sqrt{M_{J}^{2}+\dfrac{1}{\alpha^{'}}} .
\end{equation}

Using this equation we calculate the masses of excited states of $\Omega_{c}^{0}$ baryon, which are presented in Table \ref{tab:table1} and Table \ref{tab:table2} for natural and unnatural parity states, respectively. We compared our results with other theoritical predictions and experimental observations where available.

\begin{table}
	\caption{\label{tab:table1}
	Masses of excited states of $\Omega_{c}^{0}$ baryon (in GeV) in $(J,M^{2})$ plane with natural parities $P=(-1)^{J-\frac{1}{2}}$.
	}
	\begin{ruledtabular}
		\begin{tabular}{lllllll}
			States & $1^{2}S_{\frac{1}{2}}$ & $1^{2}P_{\frac{3}{2}}$ & $1^{2}D_{\frac{5}{2}}$ & $1^{2}F_{\frac{7}{2}}$ & $1^{2}G_{\frac{9}{2}}$ & $1^{2}H_{\frac{11}{2}}$  \\ 
			
			\colrule
		
			Present & 2.702 &3.049 &3.360 &3.645 &3.871 & 4.156 \\
		
			LHCb \cite{LHCb} & & 3.050\\
			Belle \cite{Belle} & &3.050 \\
			PDG \cite{PDG}  & 2.695 &3.050 & \\
		    Ref. \cite{keval2020} & 2.696 &2.968 & 3.251 &3.519 & 3.783\\
			Ref. \cite{Zalak2016} & 2.695 &3.024& 3.299&3.565 \\
		   	Ref. \cite{Ebert 2011} & 2.698 &3.054 & 3.297  &3.514 &3.705
			 &\\
		   	Ref. \cite{Robert2008}&2.718 &2.977&3.196&\\
		   	Ref. \cite{Yoshida2015}&2.731&3.033&\\
			Ref. \cite{Ebert 2008} & 2.698 &3.026&3.218 \\
			Ref. \cite{Yamaguchi2015}&2.718&3.056&3.273&\\
		
		\end{tabular}
	\end{ruledtabular}
\end{table}

\begin{table}
	\caption{\label{tab:table2}
		Masses of excited states of $\Omega_{c}^{0}$ baryon (in GeV) in $(J,M^{2})$ plane with unnatural parities $P=(-1)^{J+\frac{1}{2}}$.
	}
	\begin{ruledtabular}
		\begin{tabular}{lllllll}
			States & $1^{4}S_{\frac{3}{2}}$ & $1^{4}P_{\frac{5}{2}}$ & $1^{4}D_{\frac{7}{2}}$ & $1^{4}F_{\frac{9}{2}}$ & $1^{4}G_{\frac{11}{2}}$ & $1^{4}H_{\frac{13}{2}}$  \\ 
			
			\colrule
			
		Present & 2.772 & 3.055 &3.314 &3.555 & 3.779& 3.991 \\
		PDG\cite{PDG} & 2.765  &  &\\
		
		Ref. \cite{keval2020} & 2.766 & 2.962 & 3.241 &3.503&3.756\\
		Ref. \cite{Zalak2016} & 2.765 & 3.010 & 3.276& 3.532 \\
		Ref. \cite{Ebert 2011} & 2.776 & 3.051 & 3.283 &3.485&3.665\\
		Ref. \cite{Robert2008}&2.776&3.014&3.206&\\
		Ref. \cite{Yoshida2015}&2.779&3.057&\\
		Ref. \cite{Ebert 2008} & 2.765 & 3.022& 3.237 \\
		Ref. \cite{Yamaguchi2015}&2.766&3.014&\\
		
		\end{tabular}
	\end{ruledtabular}
\end{table}

\subsection{\label{sec:level2}Masses of $\Omega_{cc}$ and $\Omega_{ccc}$}
 The values of $\alpha^{'}_{jjq}/\alpha^{'}_{iiq}$ from Eq. (\ref{eq:8}) should be a real number. So we can write Eq. (\ref{eq:8}) as,
\begin{equation}
	\label{eq:15}
	|4M^{2}_{ijq}-M^{2}_{iiq}-M^{2}_{jjq}|\geqslant 2M_{iiq}M_{jjq} ,
\end{equation}          
this inequality relation can be simplified to,
\begin{equation}
	\label{eq:16}
	2M_{ijq} \geqslant (M_{iiq} + M_{jjq}).
\end{equation}

Many theoretical studies shows that the slopes of Regge trejectories decreases with increasing quark masses \cite{Add1,Add2,Zang2007,Brisudova2000,J.L.1986,A.B.1982}. For $m_{j}>m_{i}$, we can write $\alpha^{'}_{jjq}/\alpha^{'}_{iiq}<1$. Therefore Eq. (\ref{eq:8}) is,
\begin{equation}
	\label{eq:17}
	\begin{split}
		\dfrac{\alpha^{'}_{jjq}}{\alpha^{'}_{iiq}}=\dfrac{1}{2M^{2}_{jjq}}\times[(4M^{2}_{ijq}-M^{2}_{iiq}-M^{2}_{jjq})\\
		+\sqrt{{{(4M^{2}_{ijq}-M^{2}_{iiq}-M^{2}_{jjq}})^2}-4M^{2}_{iiq}M^{2}_{jjq}}] < 1 ,
	\end{split}
\end{equation}
the above relation gives another inequality equation as,
\begin{equation}
	\label{eq:18}
	2M^{2}_{ijq}<M^{2}_{iiq}+M^{2}_{jjq} ,
\end{equation}
so from Eqs. (\ref{eq:16}) and (\ref{eq:18}) we have,
\begin{equation}
	\label{eq:19}
	\dfrac{M_{iiq}+M_{jjq}}{2}<M_{ijq}<\sqrt{\dfrac{M^{2}_{iiq}+M^{2}_{jjq}}{2}} .   
\end{equation}

This relation can be used to calculate the upper and lower limit for baryon masses $M_{ijq}$. Further to estimate the deviation of relation (\ref{eq:19}), we introduce a parameter called $\delta$ which is used to replace the signs of inequalities to equal signs. For baryons it is denoted by $\delta^{b}_{ij,q}$ and given by,
\begin{equation}
	\label{eq:20}
	\delta^{b}_{ij,q} = M^{2}_{iiq}+M^{2}_{jjq}-2M^{2}_{ijq} ,
\end{equation}
here also $i, j$ and $q$ represents the arbitrary light or heavy quarks.

Now from Eqs. (\ref{eq:5}) and (\ref{eq:6}) we can write,
\begin{equation}
	\label{eq:21}
	a_{iiq}(0)-a_{ijq}(0)=a_{ijq}(0)-a_{jjq}(0),
\end{equation}

\begin{equation}
	\label{eq:22}
	\frac{1}{\alpha_{iiq}^{'}}-\frac{1}{\alpha_{ijq}^{'}}=	\frac{1}{\alpha_{ijq}^{'}}-\frac{1}{\alpha_{jjq}^{'}} ,
\end{equation}
based on these equations we introduce two parameters,
\begin{equation}
	\label{eq:23}
	\lambda_{x}=a_{nnn}(0)-a_{nnx}(0)  ,  \gamma_{x}=\frac{1}{\alpha^{'}_{nnx}}-\frac{1}{\alpha^{'}_{nnn}} ;
\end{equation}
where $n$ represents light nonstrange quark ($u$ or $d$) and $x$ denotes $i, j$ or $q$. From Eqs. (\ref{eq:21}-\ref{eq:23})we have,

\begin{equation}
	\label{eq:24}
		\begin{split}
	a_{ijq}(0) = a_{nnn}(0)-\lambda_{i}-\lambda_{j}-\lambda_{q} ,\\ \frac{1}{\alpha_{ijq}^{'}} = \frac{1}{\alpha_{nnn}^{'}}+\gamma_{i}+\gamma_{j}+\gamma_{q}.
	\end{split}
\end{equation}

Since in baryon multiplets for $nnn$ and $ijq$ states we can write from Eq. \ref({eq:4}),
\begin{equation}
	\label{eq:25}
	\begin{split}
	J = a_{nnn}(0) + \alpha_{nnn}^{'}M^{2}_{nnn} ,\\
    J = a_{ijq}(0) + \alpha^{'}_{ijq}M^{2}_{ijq} ,
\end{split}
\end{equation}
solving Eqs. (\ref{eq:24}) and (\ref{eq:25}) we have,
\begin{equation}
	\label{eq:26}
	M^{2}_{ijq} = (\alpha^{'}_{nnn}M^{2}_{nnn} + \lambda_{i} + \lambda_{j} + \lambda_{q})\left(\frac{1}{\alpha^{'}_{nnn}}+\gamma_{i}+\gamma_{j}+\gamma_{q}\right).
\end{equation}

Therefore from Eqs.(\ref{eq:20}) and (\ref{eq:26}), we have

\begin{widetext}
	\begin{eqnarray}
			\label{eq:27} \nonumber
\delta^{b}_{ij,q} &=& M^{2}_{iiq}+M^{2}_{jjq}-2M^{2}_{ijq}    \\ \nonumber
	&=& (\alpha^{'}_{nnn}M^{2}_{nnn} + 2\lambda_{i} + \lambda_{q})\left(\frac{1}{\alpha^{'}_{nnn}} + 2\gamma_{i} + \gamma_{q}\right) + (\alpha^{'}_{nnn}M^{2}_{nnn} + 2\lambda_{j} + \lambda_{q})\left(\frac{1}{\alpha^{'}_{nnn}} + 2\gamma_{j} + \gamma_{q}\right)  \\ \nonumber
	& -&2(\alpha^{'}_{nnn}M^{2}_{nnn} + \lambda_{i} + \lambda_{j}+ \lambda_{q})\left(\frac{1}{\alpha^{'}_{nnn}} + \gamma_{i} + \gamma_{j}+\gamma_{q}\right)\\ 
	& =& 2(\lambda_{i}-\lambda_{j})(\gamma_{i}-\gamma_{j}) .
        \end{eqnarray}
\end{widetext}

So, we can say that  $\delta^{b}_{ij,q}$ is independent of quark flavor $q$. Hence by keeping $i$ and $j$ fixed and changing the $q$ quark, using Eq. (\ref{eq:20}) we have some following relations for $\frac{3}{2}^{+}$ states :\\
(I) $i=u$, $j=s$, $q=u$, $c$
\begin{equation}
	\label{eq:28}
	\begin{split}
		\delta_{us}^{(3/2)^{+}}=M^{2}_{\Delta}+M^{2}_{\Xi^{*}}-2M^{2}_{\Sigma^{*}}\\
		=M^{2}_{\Sigma_{c}^{*}}+M^{2}_{\Omega_{c}^{*}}-2M^{2}_{\Xi_{c}^{*}};
	\end{split}
\end{equation}
(II) $i=u$, $j=c$, $q=u$ ,$s$
\begin{equation}
	\label{eq:29}
	\begin{split}
		\delta_{uc}^{(3/2)^{+}}=M^{2}_{\Delta}+M^{2}_{\Xi^{*}_{cc}}-2M^{2}_{\Sigma^{*}_{c}}\\
		=M^{2}_{\Sigma^{*}}+M^{2}_{\Omega_{cc}^{*}}-2M^{2}_{\Xi_{c}^{*}};
	\end{split}
\end{equation}
(III) $i=s$, $j=c$, $q=u$ ,$c$
\begin{equation}
	\label{eq:30}
	\begin{split}
		\delta_{sc}^{(3/2)^{+}}=M^{2}_{\Xi^{*}}+M^{2}_{\Xi^{*}_{cc}}-2M^{2}_{\Xi^{*}_{c}}\\
		=M^{2}_{\Omega_{c}^{*}}+M^{2}_{\Omega_{ccc}^{*}}-2M^{2}_{\Omega_{cc}^{*}};
	\end{split}
\end{equation}
solving Eqs. (\ref{eq:28}) and (\ref{eq:29}) we have,
\begin{equation}
	\label{eq:31}
	(M^{2}_{\Omega_{cc}^{*}}-M^{2}_{\Xi^{*}_{cc}})	+ (M^{2}_{\Xi^{*}}-M^{2}_{\Sigma^{*}}) = (M^{2}_{\Omega_{c}^{*}}-M^{2}_{\Sigma_{c}^{*}}),
\end{equation}
similarly, its corresponding relation for $\frac{1}{2}^{+}$ baryonic states is,
\begin{equation}
	\label{eq:32}
	(M^{2}_{\Omega_{cc}}-M^{2}_{\Xi_{cc}})	+ (M^{2}_{\Xi}-M^{2}_{\Sigma}) = (M^{2}_{\Omega_{c}}-M^{2}_{\Sigma_{c}}),
\end{equation}
using Eqs. (\ref{eq:11}) and (\ref{eq:32}) we have the expression,
\begin{widetext}
	\begin{equation}
		\label{eq:33}
		\begin{split}
			\left(4M^{2}_{\Xi_{c}^{+}}-M^{2}_{\Omega_{cc}^{+}}+M^{2}_{\Xi_{cc}^{++}}-2M^{2}_{\Sigma_{c}^{++}}-M^{2}_{\Xi^{0}}+M^{2}_{\Sigma^{+}}\right)
-2M_{\Sigma_{c}^{++}}\sqrt{M^{2}_{\Omega_{cc}^{+}}-M^{2}_{\Xi_{cc}^{++}}+M^{2}_{\Xi^{0}}-M^{2}_{\Sigma^{+}}+M^{2}_{\Sigma_{c}^{++}}} \\
= -\sqrt{\left(4M^{2}_{\Xi_{c}^{+}}-M^{2}_{\Omega_{cc}^{+}}+M^{2}_{\Xi_{cc}^{++}}-2M^{2}_{\Sigma_{c}^{++}}-M^{2}_{\Xi^{0}}+M^{2}_{\Sigma^{+}}\right)^{2}-4M^{2}_{\Sigma_{c}^{++}}\left(M^{2}_{\Omega_{cc}^{+}}-M^{2}_{\Xi_{cc}^{++}}+M^{2}_{\Xi^{0}}-M^{2}_{\Sigma^{+}}+M^{2}_{\Sigma_{c}^{++}}\right)} .
	   \end{split}.
   \end{equation}
\end{widetext}

Using the above expression we have calculated the ground state masses of $\Omega_{cc}^{+}$ baryon. Inserting the masses $M_{\Xi_{c}^{+}}$, $M_{\Sigma_{c}^{++}}$, $M_{\Xi_{cc}}$ and $M_{\Xi}$, $M_{\Sigma}$ from PDG\cite{PDG} and $M_{\Omega_{c}^{0}}$ (calculated above) in Eq. (\ref{eq:33}), we get $M_{\Omega_{cc}^{+}}$=3.752 GeV for $J^{P}$=$\frac{1}{2}^{+}$. Similarly we get $M_{\Omega_{cc}^{+}}$=3.816 GeV for $J^{P}$=$\frac{3}{2}^{+}$. Since the quark configuration of $\Omega_{cc}^{++}$ is \textit{scc}, hence using relation (\ref{eq:8}) we have,
\begin{equation}
	\label{eq:34}
	\begin{split}
		\dfrac{\alpha^{'}_{scc}}{\alpha^{'}_{uus}}=\dfrac{1}{2M^{2}_{\Omega_{cc}^{+}}}\times[(4M^{2}_{\Xi_{c}^{+}}-M^{2}_{\varSigma^{+}}-M^{2}_{\Omega_{cc}^{+}}) \\  +\sqrt{{{(4M^{2}_{\Xi_{c}^{+}}-M^{2}_{\varSigma^{+}}-M^{2}_{\Omega_{cc}^{+}}})^2}-4M^{2}_{\varSigma^{+}}M^{2}_{\Omega_{cc}^{+}}}] .
	\end{split}.
\end{equation}

Further using Eqs. (\ref{eq:34}) and (\ref{eq:14}) we calculate the excited state masses of $\Omega_{cc}^{+}$ baryon  in the same way we have done for $\Omega_{c}^{0}$. 
Our predicted masses are labeled with "Present" for both natural and unnatural parity states presented in Table \ref{tab:table3} and Table \ref{tab:table4} respectively, where our predictions are reasonably close to other theoretical predictions. 

\begin{table}
	\caption{\label{tab:table3}
		Masses of excited states of $\Omega_{cc}$ baryon (in GeV) in $(J,M^{2})$ with natural parities $P=(-1)^{J-\frac{1}{2}}$ .
	}
	\begin{ruledtabular}
		\begin{tabular}{lllllll}
				States & $1^{2}S_{\frac{1}{2}}$ & $1^{2}P_{\frac{3}{2}}$ & $1^{2}D_{\frac{5}{2}}$ & $1^{2}F_{\frac{7}{2}}$ & $1^{2}G_{\frac{9}{2}}$ & $1^{2}H_{\frac{11}{2}}$  \\ 
			
			\colrule
			
			Present & 3.752 &3.975 &4.186 &4.387 &4.579 & 4.763 \\
			
			Ref. \cite{ZalakC2016} & 3.650 &3.972 &4.141 &4.296 \\
			Ref. \cite{Robert2008} & 3.815 &4.052& 4.202 \\
			Ref. \cite{A.P2008} &3.719\\
			Ref. \cite{Yoshida2015} & 3.832 &4.086& 4.264 \\
			Ref. \cite{Wei2008} &3.650 & &4.174\\
			Ref. \cite{Wei2015} & 3.650 &3.910 & 4.153  &4.383 & &\\
			Ref. \cite{Ebert2002} & 3.778 &4.102 & \\
			Ref. \cite{Wang2010} &3.710 \\
			Ref. \cite{Chiu2005} & 3.637\\
			Ref. \cite{Migura2006} & 3.732 &3.986\\
			Ref. \cite{Burakovsky1997} &3.804\\
			
		\end{tabular}
	\end{ruledtabular}
\end{table}

\begin{table}
	\caption{\label{tab:table4}
	Masses of excited states of $\Omega_{cc}$ baryon (in GeV) in $(J,M^{2})$ with unnatural parities $P=(-1)^{J+\frac{1}{2}}$.
	}
	\begin{ruledtabular}
		\begin{tabular}{lllllll}
			states & $1^{4}S_{\frac{3}{2}}$ & $1^{4}P_{\frac{5}{2}}$ & $1^{4}D_{\frac{7}{2}}$ & $1^{4}F_{\frac{9}{2}}$ & $1^{4}G_{\frac{11}{2}}$ & $1^{4}H_{\frac{13}{2}}$  \\ 
			
			\colrule
			
		Present  &3.816  & 4.094 & 4.354 &4.599&4.832&5.055 \\
		
		Ref. \cite{ZalakC2016} &3.810 &3.958 &4.122 &4.274 \\
		Ref. \cite{Robert2008} & 3.876 & 4.152 & 4.230 & \\
		Ref. \cite{Brown2014} & 3.822\\
		Ref. \cite{Wei2008} & 3.808 & & 4.313 \\
		Ref. \cite{Wei2015} & 3.809 & 4.058 & 4.294 &4.516& \\

		Ref. \cite{Wang2010} & 3.760\\
		Ref. \cite{Chiu2005} & 3.762\\
		Ref. \cite{Migura2006} & 3.765\\
		Ref. \cite{Burakovsky1997} & 3.850\\
		Ref. \cite{S.S2000} & 3.730 & 4.134 & 4.204 &  \\
	
		Ref. \cite{Simonis2009} & 3.800\\
		\end{tabular}
	\end{ruledtabular}
\end{table}

Similarly for triply charmed omega baryon ($\Omega_{ccc}$), using relations (\ref{eq:30}) and (\ref{eq:33}) we have,\\
{\small
\begin{widetext}
	\begin{equation}
		\label{eq:35}
		\begin{split}
			\left(3M^{2}_{\Xi_{cc}}-M^{2}_{\Omega_{ccc}}+6M^{2}_{\Xi_{c}}-4M^{2}_{\Sigma_{c}}-M^{2}_{\Omega_{c}}+M^{2}_{\Xi}+2M^{2}_{\varSigma}\right)
			-2M_{\Sigma_{c}^{++}}\sqrt{M^{2}_{\Omega_{ccc}}-3M^{2}_{\Xi_{cc}}+M^{2}_{\Omega_{c}}+M^{2}_{\Xi}+2\left(M^{2}_{\Xi_{c}}+M^{2}_{\varSigma_{c}}-M^{2}_{\varSigma}\right)} \\
			= -\sqrt{\left(3M^{2}_{\Xi_{cc}}-M^{2}_{\Omega_{ccc}}+6M^{2}_{\Xi_{c}}-4M^{2}_{\Sigma_{c}}-M^{2}_{\Omega_{c}}+M^{2}_{\Xi}+2M^{2}_{\varSigma}\right)^{2}-8M^{2}_{\Sigma_{c}^{++}}\left[M^{2}_{\Omega_{ccc}}-3M^{2}_{\Xi_{cc}}+M^{2}_{\Omega_{c}}+M^{2}_{\Xi}+2\left(M^{2}_{\Xi_{c}}+M^{2}_{\varSigma_{c}}-M^{2}_{\varSigma}\right)\right]}  .
 		\end{split}
	\end{equation}
\end{widetext}
}
Since the ground state $(J^{P}$=$\frac{3}{2}^{+})$ mass of $\Xi_{cc}^{++}$ is not confirmed by PDG yet, so we take $M_{\Xi_{cc}}^{*}$= 3.695 GeV \cite{Wei2015} and inserting all other masses from latest PDG \cite{PDG} into Eq. (\ref{eq:35}), we obtain $M_{\Omega_{ccc}^{*}}$=4.841 GeV for $J^{P}$=$\frac{3}{2}^{+}$. Excited state masses for ${\Omega_{ccc}}$ baryon are calculated in the similar manner as we have calculated for $\Omega_{c}^{0}$ and $\Omega_{cc}^{+}$, only by changing the quark constituents i.e. $i$ ,$j$ and $q$. Our calculated masses of excited $\Omega_{ccc}$ are presented in Table \ref{tab:table5}. Our results are in accordance with other theoritical predictions.

\begin{table}
	\caption{\label{tab:table5}
		Masses of excited states of $\Omega_{ccc}$ baryon (in GeV) in $(J,M^{2})$ plane with unnatural parities $P=(-1)^{J+\frac{1}{2}}$.
	}
	\begin{ruledtabular}
		\begin{tabular}{lllllll}
			states & $1^{4}S_{\frac{3}{2}}$ & $1^{4}P_{\frac{5}{2}}$ & $1^{4}D_{\frac{7}{2}}$ & $1^{4}F_{\frac{9}{2}}$ & $1^{4}G_{\frac{11}{2}}$ & $1^{4}H_{\frac{13}{2}}$  \\
			\colrule
			
		Present & 4.841 & 5.081 & 5.310& 5.530 &5.741 &5.945 \\
	
		Ref. \cite{Robert2008}&4.965 & & 5.331\\
		Ref. \cite{A.P2008}&4.803 \\
		Ref. \cite{Wei2008}&4.818 & & 5.302 \\
		Ref. \cite{Wei2015} & 4.834& & 5.301 \\
		Ref. \cite{Migura2006}& 4.773 \\
		Ref. \cite{Burakovsky1997}&4.930 \\
		
		Ref. \cite{Jaffe1967}& 4.827\\
	    Ref. \cite{Valcare2008} & 4.758 & & 5.300\\
		Ref. \cite{Edwards2014}& 4.761 & & 5.396\\
		Ref. \cite{Rajabi2014} & 4.880\\
		\end{tabular}
	\end{ruledtabular}
\end{table}

\subsection{\label{sec:level2}Masses of $\Omega_{c}^{0}$, $\Omega_{cc}^{+}$ and $\Omega_{ccc}^{++}$ baryons in $(n,M^{2})$ Plane.}
The general equation for linear Regge trejectories in $(n,M^{2})$ plane can be expressed as,
\begin{equation}
	\label{eq:36}
	n = \beta_{0} + \beta M^{2},
\end{equation}
where $n$ = 1,2,3....is the radial principal quantum number, $\beta$ and $\beta_{0}$ are the slope and intercept of the trejectories. The Regge slope ($\beta$) and Regge intercept ($\beta_{0}$) are assumed to be same for all baryon multiplets lying on the single Regge line. So first of all these parameters ($\beta$ and $\beta_{0}$) are calculated, and with the help of them the excited states masses (\textit{$N^{2S+1}L_{J}$}, where $N$, $L$, $S$ denote the radial excited quantum number, the orbital quantum number and the intrinsic spin, respectively) of $\Omega_{c}^{0}$, $\Omega_{cc}^{+}$ and $\Omega_{ccc}^{++}$ baryons are estimated lying on these Regge trejectories. For $\Omega_{c}^{0}$ baryon, using the slope equation $\beta_{(S)} = 1/(M^{2}_{\Omega_{c}(2S)}-M^{2}_{\Omega_{c}(1S)})$, where $M_{\Omega_{c}(1S)}$=2.702 GeV (calculated above) and taking $M_{\Omega_{c}(2S)}$=3.164 GeV from \cite{Zalak2016} for $J^{P}$ = $1/2^{+}$ state, we get $\beta_{(S)}$ = 0.3689 $GeV^{-2}$. Using the following equations,
\begin{equation}
	\label{eq:37}
	\begin{split}
		1 = \beta_{0(S)} + \beta_{(S)} M^{2}_{\Omega_{c}(1S)},\\
		2 = \beta_{0(S)} + \beta_{(S)} M^{2}_{\Omega_{c}(2S)},
	\end{split}
\end{equation}
we have $\beta_{0(S)}$ = -1.6939. In the same way we can write,
\begin{equation}
	\label{eq:38}
	\begin{split}
		1 = \beta_{0(P)} + \beta_{(P)} M^{2}_{\Omega_{c}(1P)},\\
		2 = \beta_{0(P)} + \beta_{(P)} M^{2}_{\Omega_{c}(2P)},\\
		1 = \beta_{0(D)} + \beta_{(D)} M^{2}_{\Omega_{c}(1D)},\\
		2 = \beta_{0(D)} + \beta_{(D)} M^{2}_{\Omega_{c}(2D)}.
	\end{split}
\end{equation}

With the help $\beta_{(S)}$ and $\beta_{0(S)}$, we calculate the masses of excited $\Omega_{c}^{0}$ baryon for n=3,4,5... Tables \ref{tab:table6} and \ref{tab:table7} shows our calculated radial and orbital excited state masses of $\Omega_{c}$ baryon in $(n,M^{2})$ plane for natural and unnatural parity states respectively. In the similar manner we estimate the mass spectra for $\Omega_{cc}$ (see Tables \ref{tab:table8} and \ref{tab:table9}) and $\Omega_{ccc}$(see Table \ref{tab:table10}) baryons also in  $(n,M^{2})$ plane. 
\begin{table}
	\caption{\label{tab:table6}
		Masses of excited states of $\Omega_{c}^{0}$ baryon (in GeV) in $(n,M^{2})$ plane with natural parity $P=(-1)^{J-\frac{1}{2}}$. The masses from \cite{Zalak2016} are taken as input.
	}
	\begin{ruledtabular}
		\begin{tabular}{lllllll}
			\textit{$N^{2S+1}L_{J}$} & Present & \cite{Ebert 2011} & \cite{keval2020} & \cite{Robert2008} &\cite{Valcare2008} & \cite{Yoshida2015}\\
			\colrule
			
			 $1^{2}S_{\frac{1}{2}}$ & 2.702 &2.698 &2.696 &2.718& 2.699& 2.731 \\
			$2^{2}S_{\frac{1}{2}}$ & \textbf{3.164} \cite{Zalak2016} &3.088 &3.165 &3.152&3.159&3.227 \\
			$3^{2}S_{\frac{1}{2}}$ & 3.566 &3.489&3.540& & &3.292  \\
			$4^{2}S_{\frac{1}{2}}$ & 3.928 & 3.814 &3.895& \\
			$5^{2}S_{\frac{1}{2}}$ & 4.259 & 4.102 &4.238& \\
			$6^{2}S_{\frac{1}{2}}$ & 4.566 &  &4.569 \\
			\noalign{\smallskip}\hline\noalign{\smallskip}
			
			$1^{2}P_{\frac{3}{2}}$ & 3.049 &3.054 &2.968&2.986& & 3.033\\
			$2^{2}P_{\frac{3}{2}}$ & \textbf{3.408} \cite{Zalak2016} &3.433 &3.334 & & &3.056 \\
			$3^{2}P_{\frac{3}{2}}$ & 3.732 &3.752 &3.687& & &  \\
			$4^{2}P_{\frac{3}{2}}$ & 4.031 &4.036 &4.029& \\
			$5^{2}P_{\frac{3}{2}}$ & 4.309 & &4.362 \\
			\noalign{\smallskip}\hline\noalign{\smallskip}
			
			$1^{2}D_{\frac{5}{2}}$ & 3.360 &3.297 &3.251& & & \\
			$2^{2}D_{\frac{5}{2}}$ & \textbf{3.680} \cite{Zalak2016} &3.626 &3.604& & & \\
			$3^{2}D_{\frac{5}{2}}$ & 3.974 &3.752 &3.947& & &  \\
			$4^{2}D_{\frac{5}{2}}$ & 4.248 & &4.282 \\
			$5^{2}D_{\frac{5}{2}}$ & 4.505 & & \\
		\end{tabular}
	\end{ruledtabular}
\end{table}

\begin{table}
	\caption{\label{tab:table7}
		Masses of excited states of $\Omega_{c}^{0}$ baryon (in GeV) in $(n,M^{2})$ plane with unnatural parity $P=(-1)^{J+\frac{1}{2}}$. The masses from \cite{Zalak2016} are taken as input.
	}
	\begin{ruledtabular}
		\begin{tabular}{lllllll}
			\textit{$N^{2S+1}L_{J}$} & Present &\cite{Ebert 2011} & \cite{keval2020} & \cite{Robert2008} &\cite{Valcare2008} & \cite{Yoshida2015}\\
			\colrule
			
			$1^{4}S_{\frac{3}{2}}$ & 2.772 &2.768 &2.766& 2.776& 2.768&2.779 \\
		$2^{4}S_{\frac{3}{2}}$ & \textbf{3.197} \cite{Zalak2016} &3.123  &3.208&3.190&3.202&3.257 \\
		$3^{4}S_{\frac{3}{2}}$ & 3.571 &3.510&3.564& & &3.285  \\
		$4^{4}S_{\frac{3}{2}}$ & 3.910 & 3.830 &3.911 \\
		$5^{4}S_{\frac{3}{2}}$ & 4.222 & 4.114 &4.248 \\
		$6^{4}S_{\frac{3}{2}}$ & 4.513 &  & \\
		\noalign{\smallskip}\hline\noalign{\smallskip}
		
		$1^{4}P_{\frac{5}{2}}$ & 3.055 &3.051 &2.962&3.041& & 3.057\\
		$2^{4}P_{\frac{5}{2}}$ & \textbf{3.393} \cite{Zalak2016} &3.427  &3.328 & & &3.477 \\
		$3^{4}P_{\frac{5}{2}}$ & 3.700 &3.744 &3.681 & & &3.620 \\
		$4^{4}P_{\frac{5}{2}}$ & 3.983 &4.028 &4.023 \\
		$5^{4}P_{\frac{5}{2}}$ & 4.248 & & \\
		\noalign{\smallskip}\hline\noalign{\smallskip}
		
		$1^{4}D_{\frac{7}{2}}$ & 3.314 &3.283 &3.241 & & \\
		$2^{4}D_{\frac{7}{2}}$ & \textbf{3.656} \cite{Zalak2016} &3.611 &3.594 & & \\
		$3^{4}D_{\frac{7}{2}}$ & 3.968 & &3.938 & &  \\
		$4^{4}D_{\frac{7}{2}}$ & 4.258 & &4.273 \\
		$5^{4}D_{\frac{7}{2}}$ & 4.529 & & 4.594 \\
		
		\end{tabular}
	\end{ruledtabular}
\end{table}

\begin{table}[!htbp]
	\caption{\label{tab:table8}
			Masses of excited states of $\Omega_{cc}^{+}$ baryon (in GeV) in $(n,M^{2})$ plane with natural parity $P=(-1)^{J-\frac{1}{2}}$. The masses from \cite{ZalakC2016}  are taken as input.
	}
	\begin{ruledtabular}
		\begin{tabular}{lllllll}
			\textit{$N^{2S+1}L_{J}$} & Present & \cite{Ebert2002}& \cite{Yoshida2015} & \cite{Robert2008} & \cite{Wei2015}   \\ 
			\colrule
			
		$1^{2}S_{\frac{1}{2}}$ & 3.752 &3.778 &3.832 &3.815& 3.650& \\
		$2^{2}S_{\frac{1}{2}}$ & \textbf{4.028} \cite{ZalakC2016} &4.075 &4.227 &4.180& \\
		$3^{2}S_{\frac{1}{2}}$ & 4.310 &4.321&4.295& &  \\
		$4^{2}S_{\frac{1}{2}}$ & 4.564 &  & & \\
		$5^{2}S_{\frac{1}{2}}$ & 4.804 & & & \\
		$6^{2}S_{\frac{1}{2}}$ & 5.033 &   \\
		\noalign{\smallskip}\hline\noalign{\smallskip}
		
		$1^{2}P_{\frac{3}{2}}$ & 3.975 &4.102 &4.086&4.052&3.910 \\
		$2^{2}P_{\frac{3}{2}}$ & \textbf{4.259} \cite{ZalakC2016} &4.345 &4.201 &4.140 & \\
		$3^{2}P_{\frac{3}{2}}$ & 4.525 & \\
		$4^{2}P_{\frac{3}{2}}$ & 4.776 & \\
		$5^{2}P_{\frac{3}{2}}$ & 5.015 & \\
		\noalign{\smallskip}\hline\noalign{\smallskip}
		
		$1^{2}D_{\frac{5}{2}}$ & 4.186 & &4.264&4.202 &4.153 & \\
		$2^{2}D_{\frac{5}{2}}$ & \textbf{4.407} \cite{ZalakC2016} &  \\
		$3^{2}D_{\frac{5}{2}}$ & 4.617 &  \\
		$4^{2}D_{\frac{5}{2}}$ & 4.818 &  \\
		$5^{2}D_{\frac{5}{2}}$ & 5.011 &  \\
		
		\end{tabular}
	\end{ruledtabular}
\end{table}

\begin{table}[!htbp]
	\caption{\label{tab:table9}
		Masses of excited states of $\Omega_{cc}^{+}$ baryon (in GeV) in $(n,M^{2})$ plane with unnatural parity $P=(-1)^{J+\frac{1}{2}}$. The masses from \cite{ZalakC2016}  are taken as input.
	}
\begin{ruledtabular}
	\begin{tabular}{lllllll}
		\textit{$N^{2S+1}L_{J}$} & Present & \cite{Ebert2002}& \cite{Yoshida2015} & \cite{Robert2008} & \cite{Zhong2017}  \\ 
		\colrule
		
	$1^{4}S_{\frac{3}{2}}$ & 3.816 &3.872 &3.883& 3.876& 3.824& \\
	$2^{4}S_{\frac{3}{2}}$ &\textbf{4.096} \cite{ZalakC2016}&4.174&4.263&4.188&4.163 \\
	$3^{4}S_{\frac{3}{2}}$ &4.358 & &4.265&  \\
	$4^{4}S_{\frac{3}{2}}$ &4.605 &  \\
	$5^{4}S_{\frac{3}{2}}$ & 4.839 & \\
	$6^{4}S_{\frac{3}{2}}$ & 5.063 &  & \\
	\noalign{\smallskip}\hline\noalign{\smallskip}
	
	$1^{4}P_{\frac{5}{2}}$ & 4.094 & &4.220&4.152& & \\
	$2^{4}P_{\frac{5}{2}}$ & \textbf{4.247} \cite{ZalakC2016} & \\
	$3^{4}P_{\frac{5}{2}}$ & 4.394 & \\
	$4^{4}P_{\frac{5}{2}}$ & 4.537 & \\
	$5^{4}P_{\frac{5}{2}}$ & 4.676 & & \\
	\noalign{\smallskip}\hline\noalign{\smallskip}
	
	$1^{4}D_{\frac{7}{2}}$ & 4.354 & \\
	$2^{4}D_{\frac{7}{2}}$ & \textbf{4.391} \cite{ZalakC2016} & \\
	$3^{4}D_{\frac{7}{2}}$ & 4.427 &  \\
	$4^{4}D_{\frac{7}{2}}$ & 4.464\\
	$5^{4}D_{\frac{7}{2}}$ & 4.500 \\

	\end{tabular}
\end{ruledtabular}
\end{table}

\begin{table}[!htbp]
	\caption{\label{tab:table10}
			Masses of excited states of $\Omega_{ccc}^{++}$ baryon (in GeV) in $(n,M^{2})$ plane with unnatural parity $P=(-1)^{J+\frac{1}{2}}$. The masses from \cite{Zalak2017}  are taken as input.
	}
	\begin{ruledtabular}
		\begin{tabular}{lllllll}
			\textit{$N^{2S+1}L_{J}$} & Present & \cite{Robert2008}& \cite{Wei2015} &\cite{Wei2008} & \cite{Valcare2008}  \\ 
			\colrule
			
			$1^{4}S_{\frac{3}{2}}$ & 4.841 &4.965 &4.834 &4.818 &4.758\\
			$2^{4}S_{\frac{3}{2}}$ &\textbf{5.300} \cite{Zalak2017} &5.313 \\
			$3^{4}S_{\frac{3}{2}}$ &5.722  \\
			$4^{4}S_{\frac{3}{2}}$ &6.115 &  \\
			$5^{4}S_{\frac{3}{2}}$ &6.484 & \\
			$6^{4}S_{\frac{3}{2}}$ &6.834&  & \\
			\noalign{\smallskip}\hline\noalign{\smallskip}
			
			$1^{4}P_{\frac{5}{2}}$ & 5.081& & \\
			$2^{4}P_{\frac{5}{2}}$ & \textbf{5.553} \cite{Zalak2017} & \\
			$3^{4}P_{\frac{5}{2}}$ & 5.987 & \\
			$4^{4}P_{\frac{5}{2}}$ & 6.393& \\
			$5^{4}P_{\frac{5}{2}}$ & 6.774 & & \\
			\noalign{\smallskip}\hline\noalign{\smallskip}
			
			$1^{4}D_{\frac{7}{2}}$ & 5.310 &5.331 &5.301&5.302&5.300 \\
			$2^{4}D_{\frac{7}{2}}$ & \textbf{5.961} \cite{Zalak2017} & \\
			$3^{4}D_{\frac{7}{2}}$ & 6.547 &  \\
			$4^{4}D_{\frac{7}{2}}$ & 7.085\\
			$5^{4}D_{\frac{7}{2}}$ & 7.585 \\
			
		\end{tabular}
	\end{ruledtabular}
\end{table}

\section{Result and discussion}
In the framework of regge phenomenology , we have obtained the radial and orbital excited states of singly, doubly and triply charmed omega baryons. In the present work we have focused on studying the masses of $\Omega_{c}^{0}$ baryon by obtaining the mass relations derived from regge phenomenology. Ground state as well as excited state masses are obtained successfully in both $(J,M^{2})$ and $(n,M^{2})$ planes for the following baryons.  
\begin{itemize}
\item \textbf{$\Omega_{c}$ baryon :}Our calculated masses in $(J,M^{2})$ plane for $\Omega_{c}^{0}$ baryon are shown in Table \ref{tab:table1} and Table \ref{tab:table2} for natural and unnatural parity states respectively. Firstly we compared our predicted ground state masses with experimently available data \cite{LHCb,Belle,PDG} and other theoretical predictions \cite{Ebert 2011,keval2020,Zalak2016,Ebert 2008,Robert2008,Yamaguchi2015,Yoshida2015}. Our results are very close to \cite{PDG} with a slight mass difference of 7-8 MeV and also they are in good agreement with Refs.\cite{Ebert 2011,keval2020,Zalak2016,Ebert 2008}. Further the excited state masses are compared with other theoretical outcomes. For $1^{2}P_{\frac{3}{2}}$ and $1^{2}P_{\frac{5}{2}}$ states our results are in accordance with Refs.\cite{Ebert 2011,Zalak2016,Ebert 2008,Yamaguchi2015,Yoshida2015} with mass difference of range 5-50 MeV and for $1^{2}D_{\frac{5}{2}}$-$1^{2}F_{\frac{7}{2}}$ and $1^{4}D_{\frac{7}{2}}$-$1^{4}F_{\frac{9}{2}}$ states our predictions are consistent with Refs. \cite{Ebert 2011,keval2020} with mass difference of 30-65 MeV. Similarly in $(n,M^{2})$ plane, our calculated results are shown in Tables \ref{tab:table6} and \ref{tab:table7} for natural and unnatural parity states respectively. Our predicted masses are in agreement with Refs.\cite{Ebert 2011,keval2020,Robert2008}. The experimentally observed state $\Omega_{c}(3050)^{0}$ having mass 3.050 GeV is close to our prediction 3.049 GeV, so we assigned $\Omega_{c}(3050)^{0}$ as $1P$ state with $J^{P}$=$3/2^{-}$ for $S$ = $1/2$. Other four states $\Omega_{c}(3000)^{0}$, $\Omega_{c}{(3066)}^0$, $\Omega_{c}{(3090)}^0$ and $\Omega_{c}{(3119)}^0$ are not identified in this work (it may belong to remaining $1P$ states such as $1^{2}P_{\frac{1}{2}}$, $1^{4}P_{\frac{1}{2}}$ and $1^{4}P_{\frac{3}{2}}$ \cite{Rosner}). Since we get masses for $1^{2}S_{\frac{1}{2}}$, $1^{2}P_{\frac{3}{2}}$, $1^{2}D_{\frac{3}{2}}$,... and $1^{4}S_{\frac{3}{2}}$, $1^{4}P_{\frac{5}{2}}$, $1^{4}D_{\frac{7}{2}}$,...  states for natural and unnatural parity respectively in $(J,M^{2})$ plane lying on Regge line.
\end{itemize}

\begin{itemize}
\item \textbf{$\Omega_{cc}$ baryon :}
Table \ref{tab:table3} and Table \ref{tab:table4} shows  our estimated results for natural and unnatural parity states respectively in $(J,M^{2})$ plane. Our calculated ground state $(1^{2}S_{\frac{1}{2}})$ mass is in good agreement with the predictions of Refs.\cite{Robert2008,Ebert2002,Wang2010,Migura2006,Burakovsky1997,A.P2008} with mass difference of 25-65 MeV, and for $1^{2}P_{\frac{3}{2}}$-$1^{2}F_{\frac{7}{2}}$ states our results are in accordance with Refs.\cite{Wei2015,Wei2008,Robert2008,Migura2006} with mass difference of 20-70 MeV (see Table \ref{tab:table3}). Similarly for $1^{4}S_{\frac{3}{2}}$ state our predicted mass is very close to the results of Refs.\cite{Wei2015,Wei2008,Brown2014,Simonis2009} having slight mass difference of 7-16 MeV, and for $1^{4}P_{\frac{5}{2}}$-$1^{4}F_{\frac{9}{2}}$ states our results are in accordance with other theoritical outcomes (see Table \ref{tab:table4}). In the same manner we compared our calculated radial and orbital excited state masses evaluated in $(n,M^{2})$ plane for both natural and unnatural parity states and our results are consistent with other theoretical and phenomenological studies (see Tables \ref{tab:table8} and \ref{tab:table9}).
\end{itemize}
\begin{itemize}
\item \textbf{$\Omega_{ccc}$ baryon :}
For triply charmed omega baryon, we calculate the ground state and excited state masses in both $(J,M^{2})$ and $(n,M^{2})$ planes shown in Table \ref{tab:table5} and Table \ref{tab:table10} respectively. The masses the ground state $\Omega_{ccc}^{++}$ vary in the range 4.750-4.950 GeV in other theoretical references (Table \ref{tab:table5}). Our predicted mass for $(1^{4}S_{\frac{3}{2}})$ state shows few MeV difference with Refs. \cite{Wei2015,Wei2008,A.P2008,Jaffe1967}. For excited state very few results are availabe from previous theoretical predictions \cite{Robert2008,Wei2008,Wei2015,Valcare2008}, and our calculate masses are in accordance with them.
	\end{itemize}
We have successfully employed Regge phenomenological approach to calculate the masses of singly, doubly and triply charmed $\Omega$ baryons. This study will definitely help future experimental studies at LHCb, Belle II and the upcoming facility PANDA, to identify these baryonic states from resonances. 

\begin{acknowledgments}
	One of the authors Juhi Oudichhya inspired by the work of  Ke-Wei Wei, Bing Chen, Xin-Heng Guo, De-Min Li, Bing Ma, Yu-Xiao Li, Qian-Kai Yao, Hong Yu on Regge phenomenology and would like to thank them for their valuabe contributions to this field.
\end{acknowledgments}

\end{document}